\documentclass[sigconf]{acmart}

\usepackage{booktabs} 

\usepackage{multirow}
\usepackage{subcaption}
\usepackage{caption}
\usepackage{setspace}
\usepackage{amsmath}
\usepackage[english]{babel}
\definecolor{mygray}{gray}{0.5}
\usepackage{float}
\usepackage{flushend}
\usepackage{mathtools}
\usepackage{soul}
\usepackage{arydshln}            
\usepackage{breqn}
\usepackage{enumitem}
\usepackage{balance}

\usepackage{pifont}

\long\def\comment#1{}
\long\def\comments#1{}

\author{Yingrui Yang}
\affiliation{%
  \institution{Department of Computer Science, University of California}
  \city{Santa Barbara}
  \state{California}
  \postcode{93106}
  \country{USA}
}

\author{Yifan Qiao}
\affiliation{%
  \institution{Department of Computer Science, University of California}
  \city{Santa Barbara}
  \state{California}
  \postcode{93106}
  \country{USA}
}

\author{Shanxiu He}
\affiliation{%
  \institution{Department of Computer Science, University of California}
  \city{Santa Barbara}
  \state{California}
  \postcode{93106}
  \country{USA}
}

\author{Tao Yang}
\affiliation{%
  \institution{Department of Computer Science, University of California}
  \city{Santa Barbara}
  \state{California}
  \postcode{93106}
  \country{USA}
}

\renewcommand{\shortauthors}{Yingrui Yang, Yifan Qiao, Shanxiu He,  \& Tao Yang}



\AtBeginDocument{}

\title{Weighted  KL-Divergence for Document Ranking Model Refinement}

\newcommand{\cmark}{\ding{51}}%

\begin{document}

\begin{abstract}

Transformer-based retrieval and reranking models for text document search are often refined
through  knowledge distillation together  with contrastive learning.
A tight distribution matching between the teacher and student models can be hard as
over-calibration may  degrade training effectiveness when a teacher does not perform well.
This paper contrastively reweights  KL divergence terms
to prioritize the alignment between a student and a teacher model
for proper separation of positive and negative documents.
This paper analyzes and evaluates the proposed loss function  on the
MS MARCO and  BEIR datasets to demonstrate its effectiveness in improving the relevance of tested student  models.

\end{abstract}

\maketitle

\section{Introduction}

\comments{

A transformer based cross-encoder for document ranking
can achieve impressive relevance at a cost of extremely expensive computational complexity during inference.
There are many simplified models developed to reduce the time and/or space complexity of neural ranking during inference, 
including sparse retrieval with learned representations (e.g. 
SPLADE~\cite{ Formal2021SPLADE,Formal2022SPLADE++})
and a dual encoder  architecture 
with single-vector and multi-vector document representations 
(e.g. ~\cite{Khattab2020ColBERT, Santhanam2021ColBERTv2}).
}
Learned sparse representations~\cite{ Formal2021SPLADE,Formal2022SPLADE++,
Mallia2021deepimpact, Lin2021unicoil,2021NAACL-Gao-COIL}
, dual-encoder dense models  with single vector representation
(e.g. ~\cite{Lin2021tctcolbert, 2021CIKM-JPQ-Zhan, xiong2021-ANCE, gao-2021-condenser,Ren2021RocketQAv2})
and multi-vector representations 
(e.g. ~\cite{Khattab2020ColBERT, Santhanam2021ColBERTv2,
qian2022multivector, li2022citadel, li2023slim}) 
are popular neural information retrieval methods.  
These methods have been developed as simplifications of expensive cross-encoder neural ranking architectures (e.g., BERT [5]) for faster online inference.
To boost the relevance of these simplified neural models, 
knowledge distillation~\cite{Hinton2015Distilling} 
has been shown to be  critical  to transfer knowledge from a powerful teacher model during training 
through behavior imitation.
KL divergence is a  commonly used training loss for knowledge 
distillation in document 
ranking~\cite{Lin2021tctcolbert, Santhanam2021ColBERTv2,  Ren2021RocketQAv2,    Wang2022SimLM,  Liu2022RetroMAE}.

This paper explores opportunities to further improve the KL-divergence loss function based knowledge distillation in document ranking.
One limitation of KL-divergence loss in ranking
is its demand for a tight distribution match in each document, lacking prioritization. This can lead to over-calibration issues, particularly when the teacher model performs incorrectly. 
Prior research has incorporated the weighted sum of a contrastive loss with KL divergence as a regularization 
to mitigate overfitting, including a recent  study~\cite{2023ICTIR-Yang} named BKL to regularize KL divergence with  an entropy and L1-norm based loss. 
This BKL regularization still behaves incorrectly in several cases  
and our work will improve the KL divergence loss while avoiding the misbehavior of BKL.

The contribution of this paper is to propose an 
easy-to-implement modification of KL-divergence loss called contrastively-weighted KL divergence (CKL). 
Instead of following the aforementioned regularization approach,  this new formula guides knowledge distillation
in ranking by differentiating the role of positive and negative documents  in each query through weighting. 
Specifically it  
prioritizes important alignments between a student model and a teacher model and allows dynamic weight adjustment for
KL-divergence terms based on relative performance of teacher and student’s models in scoring a positive or negative document.
This paper gives a design justification on weight prioritization choices and provides an evaluation with MS MARCO and BEIR datasets
in improving student ranking models for two-stage search and dense retrieval. 

\section{Background }
\label{sect:related}

{\bf Problem definition.}
Given query $Q$, document search on  
a collection of $N$ text documents  (i.e., $\mathcal{D} = \{d_i\}^N_{i=1}$ ) finds top $k$ results, whose ranking primarily determined by query-document similarities. 
%
Let $\mathcal{D}^+$ be the subset of all  positive documents for query $Q$,
and $\mathcal{D}^-$ be  a subset containing all  negative documents for this query.
\comments{
We assume that in a training dataset, 
all positive documents are ranked equally  and the rank positions of negative  documents may not be important
as long as they are ranked below positive documents for a query.
}
The top one  probability distribution over these documents is defined as: 
 $
    P(d_i|Q, \mathcal{D}^+, \mathcal{D}^-, \Theta) = \frac{\exp(S(Q,d_i,\Theta))}{\sum_{j=1}^{N}\exp(S(Q,d_j,\Theta))}
$,
where $\Theta$ is the vector of neural parameters involved. $S(Q,d_i,\Theta)$ is a scoring function that captures
the semantic similarity of a document with query.
\comments{
For the simplicity  of presentation when no confusion is caused, 
we will not list $\Theta$ and $Q$ explicitly in each symbol below
and the loss function is specified for each query $Q$ based on parameters $\Theta$ under
the training documents $\mathcal{D}^+$ and $\mathcal{D}^-$.  
}
Knowledge distillation is  a training methodology  that guides the refinement of a  neural student  model using a teacher model.
Let $p_i$ or $q_i$ denote $P(d_i|Q, \mathcal{D}^+, \mathcal{D}^-, \Theta)$ where $p_i$ and $q_i$  
refer to the teacher's and student's prediction, respectively. 


To train a ranking model, the standard loss  function includes  the negative log likelihood or its variation:
$-\sum_{d_j \in \mathcal{D}^+}  \log q_j.$ 
KL-divergence defined below 
measures a distance between teacher's and student's distributions, and
has  become  a dominating choice for knowledge distillation for ranking model refinement
as  seen in the recent 
studies~\cite{Wang2022SimLM,Shen2022unifiedretriever, Zhang2022AR2, Ren2021RocketQAv2,Santhanam2021ColBERTv2}. 
$
    \it{L}_{KL} =  
\sum_{d_i \in  \mathcal{D}^+ \cup \mathcal{D}^- } p_i \ln\frac{p_i}{q_i}.
$

KL divergence does not differentiate  the role of  positive and negative documents in a training query 
and  forces  a student model to closely match the teacher scoring  in all documents.
In practice, it is not easy for a student to imitate the behavior of a teach model perfectly in all cases. 
When  a teacher does not perform as good as a student, 
KL divergence still accumulates the discrepancy between the student and the teacher as a loss without 
detecting and deprioritizing such a case.
Thus
the  key weakness of knowledge distillation is that it lacks a prioritization and can lead to over calibration.
While the BKL loss~\cite{2023ICTIR-Yang} improves this, 
its regularization formula over-corrects the behavior of KL divergence 
and  incorrectly lets  the student model deviate from or follow teacher's ranking score in a wrong learning direction
for several important cases as discussed in Section~\ref{sect:justify} (Figure~\ref{fig:grad}).
\comments{
A common approach to balance knowledge distillation is to  combine  the KL divergence loss with a contrastive loss using a weighted sum as a regularization.
While a contrastive loss for ranking often uses the log-likelihood, 
the contrastive loss proposed in BKL~\cite{2023ICTIR-Yang} 
combines the negative entropy component of positive documents and the L1-norm expression of negative documents for a given query
to balance knowledge distillation.
The weakness of BKL is that its formula over-corrects the behavior of KL divergence in some important cases.
When a teacher's model performs much better than a student in ranking a negative example, BKL's regularization 
formula incorrectly lets  the student model deviate from teacher's ranking score in a wrong learning direction.
}


\section{Method Proposed} 
\label{sect:CKL}

\subsection{Contrastively-weighted KL-divergence}

\comments{
KL divergence 
does not differentiate  the role of  positive and negative documents for a training query. 
When  a teacher does not perform as good as a student, KL divergence does not detect such a case
and still forces a student to closely match scoring  of the teacher model.  
}

To address the above limitation of KL divergence loss, 
our design introduces a weight for each 
divergence term $p_i \log\frac{p_i}{q_i}$ 
to explicitly prioritize 
the separation of positive and negative documents through weight adjustment. 
We down-weight positive documents 
ranked high on the top positions, and negative documents ranked low
at the bottom positions by a student model.
Specifically for a positive document $d_j$, the goal is to have student score $q_j$ as large as possible towards 1, and 
thus we use $(1-q_j)^\gamma$ as the weight.
Here $\gamma$ is a fixed exponent hyperparameter controlling the scale of weight and we 
set $\gamma \ge 1$.
For a negative document $d_i$, the goal is to have student score  $q_i$ as small  as possible towards 0, and 
we use $(q_i)^{\gamma-\beta_i}$ as the weight.
The bias term $\beta_i$ in the exponent adds another control to fine-tune the weight 
for a negative document.
We require $\gamma -\beta_i \ge 1$.

The  proposed {\em contrastively-weighted} KL-divergence (CKL) is:
\begin{equation*}
\begin{aligned}
    \it{L}_{CKL} = & \sum_{ d_j \in \mathcal{D}^{+}} (1-q_j)^{\gamma} p_j \ln\frac{p_j}{q_j} 
     + \sum_{d_i \in \mathcal{D}^{-}} (q_i)^{\gamma - \beta_i } p_i \ln\frac{ p_i }{q_i}.
\end{aligned}
\end{equation*}

For negative document $d_i$, exponent weight bias
\[
\beta_i= 
\alpha \left(
\frac{1}{\pi(i)}
- \frac{ 1}{ | \mathcal{D}^+|}   \sum_{d_j \in \mathcal{D}^{+}} \frac{1}{\pi(j)}
\right).
\]
Here
$\pi(i)$, $\pi(j)$ are the rank of negative document $d_i$ and positive document $d_j$ respectively.
Bias $\beta_i$ represents the importance of correcting the ranking position of
negative document $d_{i}$, compared against  the harmonic average position of positive documents.
The above use of  a rank position  is motivated by the previous work which considers 
the relevance gain by swapping two documents in a ranked order, e.g.
LambdaMART~\cite{burges2010ranknet} and 
CL-DRD ~\cite{Zeng2022Curriculum}. 

\begin{figure}[htbp]
    \centering
    \includegraphics[scale=0.75]{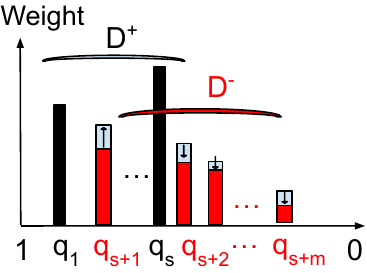}
    \caption{The weights of CKL terms, 
sorted in a descending  order of student's predictions} 
    \label{fig:KLweight}
\vspace{-1em}
\end{figure}

Figure~\ref{fig:KLweight} depicts the weight of KL-divergence terms where 
$x$ axis lists documents in a descending order of their top one probability. 
Let $s= | \mathcal{D}^+|$ and  $m= | \mathcal{D}^-|$. 
Without loss of generality, 
let  $\mathcal{D}^+=\{d_1, \cdots, d_s\}$, and $\mathcal{D}^- =\{d_{s+1}, \cdots, d_{s+m}\}$,
and let  $q_j \ge q_{j+1}$ for all positive examples $1\leq j<s-1$
and $q_i \ge q_{i+1}$ for all negative examples $s+1\leq i<s+m$.

\begin{itemize}[leftmargin=*]
\item 
For all positive documents $d_j $ where $1\le j <s$, 
 because 
$
(1-q_j)^{\gamma} \le (1-q_{j+1})^{\gamma},  
$ 
%
a low-scoring positive document is weighted  more than
a high-scoring positive document, which represents our design priority for  
the  boundary separation of positive and negative documents.


\item  Among negative documents, assuming $q_{i} > q_{i+1}$, document $d_i$ is ranked before $d_{i+1}$.
$ \frac{1} {\pi(i)}  >  \frac{1} {\pi(i+1)}$.
Thus 
$\beta_i > \beta_{i+1}$. We have 
$
(q_i)^{\gamma-\beta_i}
\ge (q_i)^{\gamma-\beta_{i+1}}
  \ge (q_{i+1})^{\gamma-\beta_{i+1}}. 
$
High-scoring negative documents are weighted more.




\item   
For any positive and negative  document pair $d_j$ and  $d_i$ where $1\leq j \leq s$ and $ s+1 \leq i \leq s+m$,
we discuss their relative weight. 
\begin{itemize}[leftmargin=*]
\item When $\beta_i \leq  0$,
this negative document $d_i$ is ranked lower than the harmonic average position of  positive documents.
The degree of importance to learn from the teacher in scoring alignment for this document decreases. Thus
$
(1-q_j)^{\gamma} \ge  
(q_{i})^{\gamma-\beta_i}.
$
Notice $1-q_j \ge q_i$ and $\gamma \le \gamma - \beta_i$.
\comments{
Therefore CKL ensures that positive document $d_j$ gains more priority in its  scoring alignment with the teacher's model
than negative document $d_i$.
}
\item When $\beta_i > 0$,  
negative document $d_i$ is ranked too high compared to the harmonic average position of  positive documents.
Correcting the  rank of this negative document becomes important.
%
CKL upweights the corresponding negative document, and narrows its 
priority gap to a positive document.
The light blue area in each weight bar of a negative document in Figure \ref{fig:KLweight} depicts such an effect with up or down
arrows.
\end{itemize}
\end{itemize}

\subsection {Additional Justifications }
\label{sect:justify}

We can prove the following properties of CKL.
\begin{equation}
\label{eq:CKLbound}
\begin{aligned}
\it{L}_{CKL} \ge & \sum_{d_j \in \mathcal{D}^+\bigcup \mathcal{D}^{-} 
} p_j \ln \frac{p_j} {q_j}  
 +  \sum_{ d_i  \in \mathcal{D}^{-}} p_i (1- q_i^{\gamma-\beta_i})  \ln q_i  \\
&
+\frac{\gamma}{\log e} \sum_{d_j \in \mathcal{D}^{+} }  q_j \log q_j \\
\end{aligned}
\end{equation}
The first component of the right hand side (RHS) is KL divergence. The third component of RHS is negative entropy of positive documents.
The sum of the first and second components in RHS approaches a constant lower bound, reached
when  $p_i=q_i$ for all positive documents and  $q_i=0$ for all negative documents. 
The third component of RHS is bounded  by $-\frac{2\gamma}{e}$, approached when all   $q_j$ values  are equal for all positive documents $d_j$. 
Thus CKL has a constant lower bound below where $p_i$ values from the teacher's model are a constant.  
$$
\textit{L}_{CKL} \ge \sum_{d_i \in \mathcal{D}^-} p_i (-1+ \ln p_i) -\frac{2\gamma}{e}.
$$
\begin{itemize}[leftmargin=*]
\item {\bf Loss minimization relationship with KL and entropy}.  
This result shows that minimizing CKL inherently provides a tradeoff between minimizing KL 
divergence, maximizing entropy of positive documents (does not favor particular positive documents over others) 
while favoring low scores of all negative documents.

\item {\bf Bounded training target.}
The  above result shows CKL loss has a constant lower bound. 
If unbounded, training would be unstable and hard to converge.     
This property indicates CKL preserves the bounded nature as KL-Divergence which
has  a lower  bound 0 when  $p_i=q_i$ for all documents  $d_i$.



\end{itemize}

 

{\bf Relative gradient contributions}.
\label{sect:gradient}
Let $L$ be the loss function $L$ where $L$ can be $L_{KL}$, $L_{CKL}$ or $L_{BKL}$. $L(i)$ is the loss corresponds to document $i$.
Let $\theta$ be a parameter used in the neural network that maps
the input features to score $S(Q,d_i,\Theta)$ for each document $d_i$ under loss $L$. Then
\begin{small}
\begin{dmath*}
\frac{\partial L} {\partial \theta} = 
\sum_{d_i \in \mathcal{D}^+ \bigcup \mathcal{D}^-} \frac{\partial L(i) } {\partial  q_i }
\frac{\partial q_i} {\partial  S(Q,d_i,\Theta)}
\frac{\partial S(Q,d_i,\Theta) }   {\partial \theta},
\end{dmath*}
\end{small}

where $q_i$ is the student score of document $i$. We examine the gradient contribution 
$\frac{\partial L(i) } {\partial  q_i} $  from document $d_i$ in computing
$\frac{\partial L } {\partial \theta}$ during the SGD-based training.  
To understand the relative gradient ratio between $\frac{\partial L_{CKL} } {\partial  \theta} $ and
$\frac{\partial L_{KL} } {\partial  \theta} $, we compare the document-level  pairwise ratio of the gradient  contribution between CKL and KL losses.
$
g_{CKL} = \frac{\partial \it{L}_{CKL}(i)}{\partial q_i} / \frac{\partial \it{L}_{KL}(i)}{\partial q_i}.
$
\begin{figure}[h!]
    \centering
    \includegraphics[scale=0.33]{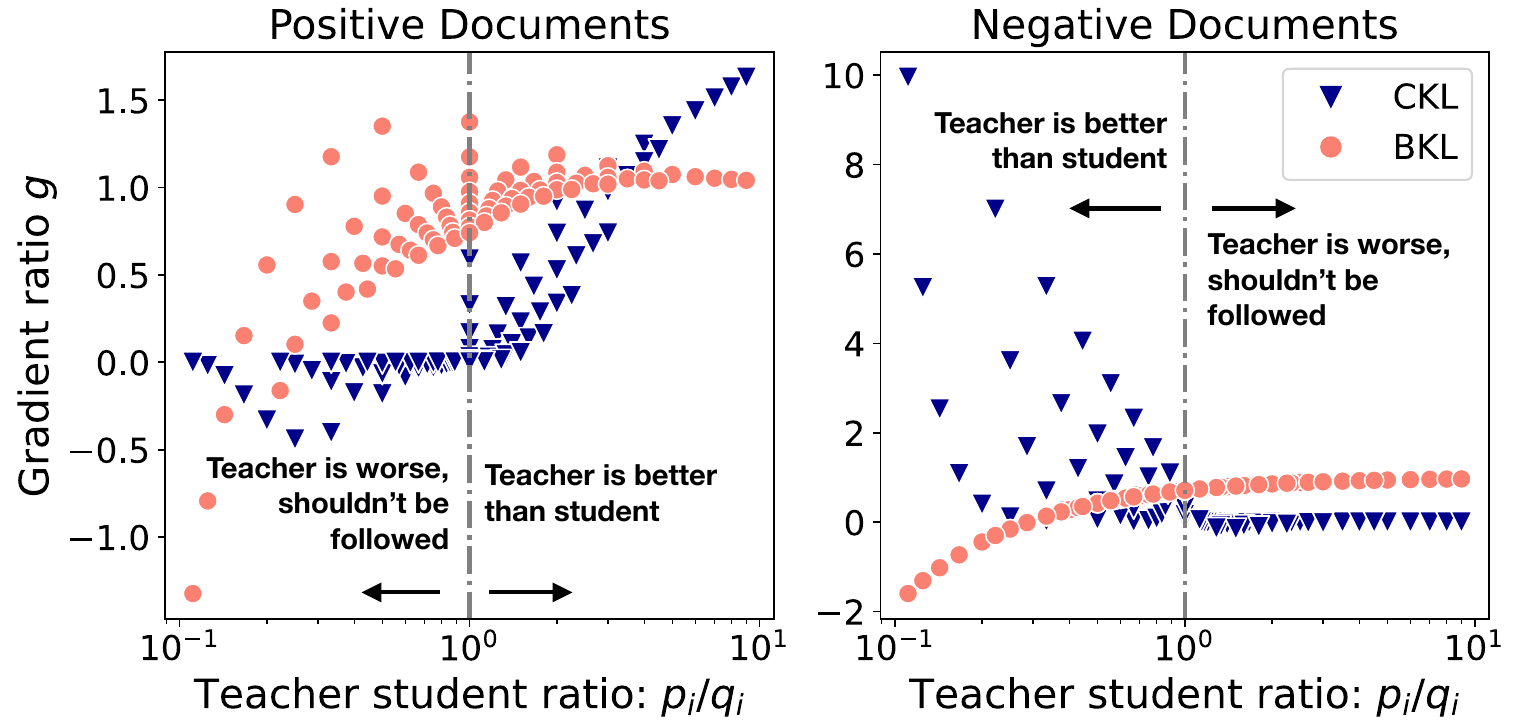} 
\vspace{-1em} 
    \caption{Relative gradient contribution ratio $g$ of CKL in blue triangle and BKL in red bullets}
    \label{fig:grad}
\vspace{-2mm}
\end{figure}

Figure~\ref{fig:grad} plots the value of gradient contribution ratio $g_{CKL}$ for CKL ($g_{CKL}$) in  a blue triangle,
when $p_i$ and $q_i$ vary from 0 to 1. The x-axis is the ratio of teacher student prediction $\frac{p_i}{q_i}$. This figure also depicts relative gradient contribution ratio $g_{BKL} $ of BKL’s formula~\cite{2023ICTIR-Yang} in a red bullet:
$
 g_{BKL} = \frac{\partial \it{L}_{BKL}(i)}{\partial q_i} / \frac{\partial \it{L}_{KL}(i)}{\partial q_i}.
$ We fix $\gamma=5$ and $\beta_i=0$ in this example for CKL and $\lambda=0.1$ for BKL for visualization purpose. 

Figure~\ref{fig:grad} illustrates that for positive documents, when the teacher performs better than the student with $\frac{p_i}{q_i} > 1$,
$g_{CKL}>0$ or  exceeds 1 and CKL allows the student to follow the teacher’s parameter update direction.
When the teacher underperforms with $\frac{p_i}{q_i} < 1$,
$g_{CKL}$ become close to 0 or even negative, and the  student does not learn much from the teacher or its learning  
deviates from teacher's learning direction. 
In comparison, BKL still forces the student to follow the teacher's direction with $g_{BKL} >0$ or even $>1$  in most of cases
when the teacher is worse.

Similarly for negative documents, when the teacher is better with $\frac{p_i}{q_i} <1$, $g_{CKL}>0$ and the student  follows  the teacher. 
But $g_{BKL}$ is close to zero and can be negative, meaning the student model does not follow and even incorrectly deviates from teacher. 
When the teacher is worse with $\frac{p_i}{q_i} > 1$, $g_{CKL}$ is close to 0 and the student does not learn much from the teacher.
But in this case, $g_{BKL}>0$, varying up to 1 and   meaning  that the student model still follows the teacher, even conservatively.

In summary, CKL’s design corrects the misbehavior of BKL by directly weighting KL divergence terms.

{\bf Discussion on $\beta_i$}.
Exponent bias $\beta_i$  for negative document $d_i$ in CKL is updated based on its rank position
immediately after each training iteration where
$q_i$ is recomputed, which makes   the loss function non-differentiable.
Thus  during training, we opt to periodically update $\beta_i$ using the latest student's model performance,
and the priority adjustment of each negative document is  stable for a block of training iterations.
This design  allows $\beta_i$ to be treated as a constant in the loss function.
This is a reasonable tradeoff as model refinement that addresses ranking accuracy for a negative document
takes a number of iterations and continuous $\beta_i$  adjustment for such a document may not yield sufficient benefits.
Hyperparameter $\alpha$ controls the maximum $\beta_i$ value. We set $0\leq \alpha \leq \gamma -1$ which ensures that $\gamma-\beta_i \ge 1$.
It is easy to  verify that
$
         |\beta_i|  \leq  \alpha( 1- \frac{ 1}{ |\mathcal{D}^+ \cup \mathcal{D}^-|} ) <  \alpha.
$

\vspace{-4mm}
\section{Evaluation Results}
\label{sect:evaluation}

{\bf Evaluation setup}.
To evaluate in domain performance, we use the MS MARCO dataset with 8.8 million passages~\cite{Campos2016MSMARCO,Craswell2020OverviewOT}. 
The test query sets include
Dev with 6980 queries, and  TREC DL 2019/2020 tracks with 43 and 54 queries.
\comments{
we report mean reciprocal rank (MRR@10) for relevance instead of 
using normalized discounted cumulative gain (NDCG)~\cite{NDCG}  because  
such a set has about one judgment label per query, which is too sparse to use NDCG.
For TREC DL test sets which have many judgment labels per query, we report the commonly used NDCG@10 score.
When space allows, we also report the recall ratio at 1000 which is 
the percentage of relevant-labeled results appeared in the final top-1000 results.
}
 To assess the zero-shot performance of a trained student model, 
we use BEIR which contains 13 publicly available datasets~\cite{thakur2021beir}, including
DBPedia, FiQA, NQ, HotpotQA, NFCorpus, T-COVID, Touche, ArguAna, C-Fever, Fever, Quora, Scidocs, and SciFact.
CKL is applied to refine the following two student models: 
1) A two-stage search pipeline that combines 
ColBERT re-ranking  with a multi-vector representation~\cite{Santhanam2021ColBERTv2}
and Sparse first-stage SPLADE retriever with a learned neural representation~\cite{Formal2022SPLADE++}. 
2) Dense single-vector retriever SimLM~\cite{Wang2022SimLM}. 
Two teachers are used: 1) MiniLM-l-6-v2~\cite{msmarcoMiniLMv2} with 0.407 MRR@10 on MS MARCO Dev on top of SPLADE retrieval 
2) a cross encoder from SimLM project~\cite{Wang2022SimLM} with 0.438 MRR@10. 

{\bf Training steps and configurations.}
\comments{
Training for each student model involves two steps: Step 1 is to 
warm up the student  model  with knowledge distillation following a fixed teacher model.
Step 2 is to use the proposed CKL loss or other loss options 
to refine the student retriever model and the student re-ranker model separately.
When we compare different loss functions for  the refinement, we 
always  start from the same model after warm-up and refine it
using  the same set of training triplets and the same teacher model. 
In this way, we rule out the potential influence caused by different implementation details in performance comparison. 
}
We start training from the officially released pretrained checkpoints. For SPLADE and ColBERT, we follow some of the warm-up settings in SPLADE++~\cite{Formal2022SPLADE++} and ColBERTv2~\cite{Santhanam2021ColBERTv2}.
The cross encoder teacher adopted is 
MiniLM-l-6-v2~\cite{msmarcoMiniLMv2}, which has been used by  
ColBERTv2 as its teacher.  
\comments{
Following the setting of SPLADE++, we 
use co-Condenser~\cite{coCondenserMarco} 
as the pretrained starting checkpoint 
and  adopt 
sentenceBERT~\cite{msmarcoHardNegatives} as
the ranker to select hard negatives. 
To train retrieval model,  we also add  additional sparsity regularization  with coefficients 0.008 and 0.01 for a query and documents
respectively, following SPLADE++.  
This observation aligns with the results reported in TAS-B~\cite{Hofsttter2021TASB}. 
}
The warm-up step uses margin-MSE~\cite{Hofsttter2020marginMSE} as the loss for knowledge distillation for both SPLADE and ColBERT. 
The above warm-up step allows 
the pipeline to deliver 0.399 MRR@10. 
For model refinement, we use the CKL loss for knowledge distillation or another loss function to compare.  
We index the corpus 
using PISA~\cite{mallia2019pisa}.  For SimLM, we compare KL Divergence, CKL or warm up with KL followed by CKL refinement. The teacher we use is the teacher released by the SimLM paper.
We use up to four NVIDIA V100 GPUs for model training.
Learning rates 2e-5 and 1e-5 are used in the warm-up step and the refinement step, respectively.  
\comments{
We update the exponent weight bias $\beta_i$ discussed in Section~\ref{sect:CKL}
every 2000 training batches, as more frequent update does not lead to an improvement. 
}
When training the student retriever, to avoid  the expensive re-indexing time during this update,
we re-evaluate the top 50 documents per training query as an approximation using the model checkpoint saved after every 2000 batches.
The above refinement with CKL for training takes less than 5 epochs to converge. 

We use $(\gamma, \alpha)=(5,1)$ for the ColBERT, SPLADE pipeline and (1,0) for SimLM. 
When comparing different loss options during training, we maintain the same setup in terms of negative samples, the initial warm-up checkpoint, and the machine environment. For the results presented in this section, we conduct paired t-tests at the 95\% confidence level. We denote results that show statistically significant degradation from CKL with `$^{\dag}$'.  We do not perform t-tests on DL’19 and DL’20 as these sets are relatively small.




\begin{table}[htbp]
	\centering
	\resizebox{0.9\columnwidth}{!}{
		\begin{tabular}{ r r r r r}
			\hline\hline
			 &\textbf{Dev}& \textbf{DL19}& \textbf{DL20}& \textbf{BEIR(Avg)}\\
			& MRR@10 &  NDCG@10 & NDCG@10 & NDCG@10 \\
\hline
SPLADE++~\cite{Formal2022SPLADE++} & 0.380  &  0.732 & -- &  0.507 \\
ColBERTv2& 0.397  &   -- & -- & 0.499 \\
\hline
\multicolumn{5}{c}{\textbf{	SPLADE + top-1000 ColBERT re-ranking }}\\
\hline
			KLDiv & 0.406$^{\dagger}$ & 0.716 & 0.719 & 0.489 \\
			MarginMSE & 0.406$^{\dagger}$ &  0.704 & 0.710 & 0.503\\
			KLDiv\_logL  & 0.405$^{\dagger}$ & 0.711 & 0.699 & 0.499\\
			CL-DRD
			& 0.406$^{\dagger}$ & 0.700 & 0.693 & 0.497\\
                BKL & 0.407 & 0.716 & 0.736 & 0.506 \\
                \hline
			\textbf{CKL}& {\bf 0.411\ } & {\bf 0.744} & {\bf 0.741} & {\bf 0.515} \\
			\hline\hline
		\end{tabular}
	}
	\caption{Two-stage search with  different loss options
}
	\label{tab:loss_pipeline}
\vspace{-4mm}
\end{table}

{\bf Two-stage student model SPLADE/ColBERT.}
Table~\ref{tab:loss_pipeline} compares the two-stage search trained under CKL and other distillation loss options
in terms of MRR@10 or NDCG@10. 
The column for BEIR lists  the average NDCG@10 across 13 datasets.
``KLDiv\_logL'' is negative log likelihood loss on in-batch negatives plus KL-divergence loss. 
Other losses include MarginMSE loss~\cite{Hofsttter2020marginMSE}, CL-DRD~\cite{Zeng2022Curriculum},  and BKL~\cite{2023ICTIR-Yang}. 
CKL visibly outperforms other loss options for MS MARCO passages, 
and also for BEIR zero-shot performance.
Overall speaking, CKL delivers a good relevance across  the tested datasets.
This table  also listed published SPLADE++  and ColBERT performance as a reference. 
\comments{
To demonstrate our evaluation is conducted competitive to the state-of-the-art research with multi-vector representations,
this table lists dense retrievers with multi-vector representations like CITADEL and ALIGNER. We also include SLIM+ which improves  multi-vector representations with a sparse scheme. The table also lists
ColBERT re-ranking with uniCOIL first-stage retrieval and CQ quantization~\cite{Yang2022CQ}.
To demonstrate that CKL performance is stronger and  competitive to the state-of-the-art research 
with multi-vector representations, this table lists dense retrievers with multi-vector representations with ALIGNER. 
These include  
ALIGNER~\cite{qian2022multivector}, 
CITADEL~\cite{li2022citadel}, and SLIM++~\cite{li2023slim} 
which obtain
MRR@10  0.403, 0.399 and 0.403 for MS MARCO Dev set and average NDCG@10 0.511, 0.501 and 0.490 for 13 BEIR datasets, 
respectively. 
}
Notice that CKL's performance number for MS MARCO and BEIR exceeds  or  is competitive to several state-of-the-art research  studies 
on  multi-vector representations~\cite{ qian2022multivector, li2022citadel, li2023slim}.

\begin{table}[h]
\resizebox{0.8\columnwidth}{!}{
        \centering
\begin{tabular}{r r r r  }
                        \hline
                        \hline Model & \textbf{Dev}& \textbf{DL19}& \textbf{DL20}\\
            & MRR@10 & NDCG@10 & NDCG@10 \\
                        \hline
\hline
SimLM w/o title anno. & 0.344 & 0.650   & 0.641  \\
               \hline
Trained with KL& 0.365 & 0.685 &  0.611 \\
Trained  with CKL& 0.381 & 0.690 &  0.696 \\
Trained with  KL+CKL& 0.391 & 0.708 &  0.706 \\
                        \hline
                        \hline
                \end{tabular}
}

        \caption{Refine dense retriever SimLM with  CKL}
\label{tab:SimLM}
\vspace{-8mm}
\end{table}

{\bf  Student dense retrieval model  SimLM}. 
CKL is applied to train on a SOTA dense retrieval model SimLM~\cite{Wang2022SimLM}
and Table~\ref{tab:SimLM} demonstrates the usefulness of CKL in SimLM.  
CKL delivers 0.391 in MRR@10 with a warmup using KL divergence. Without warmup, CKL delivers 0.381.
For dense retrievers, the released SimLM checkpoint ~\cite{SimLMCheckPoint}
gives 0.344 MRR@10 using the standard MS MARCO.
This is below 0.411 reported in ~\cite{SimLMCheckPoint}
which  evaluates on the modified MS MARCO dataset with title annotation.
Title annotation is considered unfair in ~\cite{2023SIGIR-Lassance}
since the original dataset released doesn’t utilize title information. 
The numbers reported from recent papers RocketQAv2~\cite{Ren2021RocketQAv2},
LexMAE ~\cite{shen2023lexmae}, RetroMAE and RetroMAE-2~\cite{Liu2022RetroMAE,liu-etal-2023-retromae}
were boosted by this title annotation. 
All experiments for CKL  follow the standard approach to use 
the original MS MARCO without title annotation,
and the CKL improvement in refining  SimLM is reasonable compared to KL.

\begin{figure}[htbp]
    \centering  
    \includegraphics[scale=0.32, trim=0 10cm 0 0, clip]{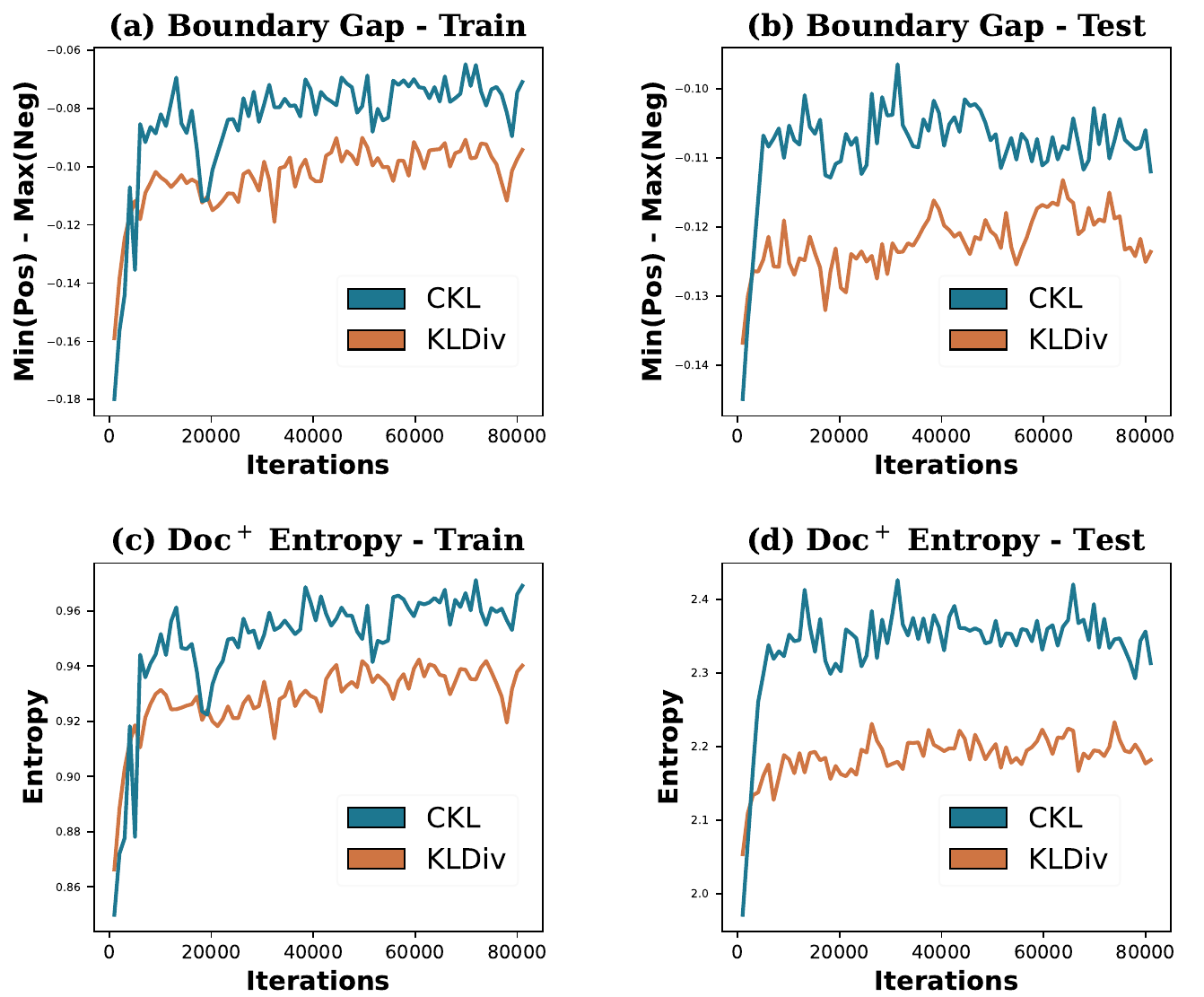}
    \caption{           
Behavior characteristics of CKL/KL during training 
}                
    \label{fig:decomp}  
\vspace{-4mm}
\end{figure}

{\bf Behavior characteristics of CKL}.
Figure~\ref{fig:decomp} shows several behavior characteristics and differences when using CKL and KL during model refinement
of two-stage search with ColBERT/SPLADE,
which reflects  our design consideration explained in Section~\ref{sect:CKL}. 
Figures~\ref{fig:decomp}(a) and (b) depict
the difference between the lowest top one probability of positive documents
with the highest top one probability  of negative documents in top 10 results predicted by a student model
during training and DL19/DL20 testing.
This result reflects the design objective of CKL loss which tries to separate positive documents from
negative documents in terms of probability distribution in each query.  
Figures~\ref{fig:decomp}(c) and (d)  show Shannon's entropy among positive documents during training
and during DL19 and DL20 testing. 
This entropy  value with CKL is higher than that of KL,
reflecting  the result of Expression~(\ref{eq:CKLbound}): minimizing CKL implies that the entropy among positive documents is maximized.

\section{Concluding Remarks}
The novelty of this work is an easy-to-implement and effective loss modification of KL-divergence for ranking model
refinement with a justification. 
The applicability of CKL is restricted to training where two-level positive and negative labels are available per query. 
That is common in practice because  it is costly to build a large training dataset for ranking with multi-level labels. 
The original MS MARCO training dataset does not contain negative documents labeled, and  
negatives are added algorithmically~\cite{xiong2021-ANCE, Karpukhin2020DPR,Ren2021RocketQAv2}.
The evaluation shows CKL can  effectively boost the relevance of tested student models,
and  achieve reasonably strong relevance numbers compared to other recent ranking studies.

{\bf Acknowledgments}. We thank anonymous referees for their valuable comments.
This work is supported in part by NSF IIS-2225942
and has used the computing resource of the ACCESS program supported by NSF.
Any opinions, findings, conclusions or recommendations expressed in this material
are those of the authors and do not necessarily reflect the views of the NSF.

\normalsize
\bibliographystyle{ACM-Reference-Format}
\balance
\bibliography{main}

\end{document}